\documentclass[sigconf]{acmart}

\AtBeginDocument{%
  \providecommand\BibTeX{{%
    \normalfont B\kern-0.5em{\scshape i\kern-0.25em b}\kern-0.8em\TeX}}}

\copyrightyear{2024}
\acmYear{2024}
\setcopyright{acmlicensed}\acmConference[GUIDE-AI '24]{Governance, Understanding and Integration of Data for Effective and Responsible AI}{June 14, 2024}{Santiago, AA, Chile}
\acmBooktitle{Governance, Understanding and Integration of Data for Effective and Responsible AI (GUIDE-AI '24), June 14, 2024, Santiago, AA, Chile}
\acmDOI{10.1145/3665601.3669844}
\acmISBN{979-8-4007-0694-3/24/06}

\usepackage{graphicx} 
\usepackage{subcaption} 
\usepackage{enumitem}
\usepackage{cleveref}
\usepackage{fancyvrb}
\usepackage{xcolor}
\usepackage{caption}

\newtheorem{theo}{Theorem}
\newtheorem{example}[theo]{Example}

\newcommand{\stitle}[1]{\vspace{2pt}\noindent\textbf{#1}}

\begin{document}

\title{Disambiguate Entity Matching using Large Language Models through Relation Discovery}





\author{Zezhou Huang}
\email{zh2408@columbia.edu}
\affiliation{
  \institution{Columbia University}
  \country{USA}
}





\begin{abstract}

Entity matching is a critical problem in data integration, central to tasks like fuzzy joins for tuple enrichment. Traditional approaches have focused on overcoming fuzzy term representations through methods such as edit distance, Jaccard similarity, and more recently, embeddings and deep neural networks, including advancements from large language models (LLMs) like GPT. However, when integrating with external databases, the core challenge in entity matching extends beyond term fuzziness to the ambiguity in defining what constitutes a "match". This is because external databases contain tuples with varying levels of detail and granularity among entities, and an "exact match" in traditional entity matching rarely happens. As a result, understanding how entities are related and the potential nuances is critical, especially for high-stake tasks for responsible AI. In this work, we study a case problem of entity matching for ESG reporting. We propose a novel approach that shifts focus from purely identifying semantic similarities to understanding and defining the "relations" between entities for resolving ambiguities in matching, with a human-in-the-loop process to make the final decision. By pre-defining a set of relations relevant to the task at hand, our method allows analysts to navigate the spectrum of similarity more effectively, from exact matches to conceptually related entities, and responsibly perform downstream tasks.

\end{abstract}

\maketitle

\section{Introduction}

Entity matching, also known as record linkage, is the fundamental task for performing fuzzy join for data integration~\cite{chen2019customizable} and deduplication for data cleaning~\cite{manghi2020entity}. 
Previous efforts concentrated on the {\it fuzzy term representations, such as synonyms and abbreviations}.
To match fuzzy terms, traditional methods like edit distance and Jaccard similarity have been used to measure term similarity. 
To capture semantic similarities between terms, techniques such as embeddings~\cite{lu2021entity} and deep neural networks~\cite{barlaug2021neural,mudgal2018deep,li2023effective} have been utilized. Recently, large language models (LLMs) like GPT have achieved results that are comparable, or even better than, previous SOTA approaches, with zero or few shot learning prompts~\cite{nuntachit2023check,narayan2022can}.

With the growing access to open data sources and data markets, there has been a higher demand for {\it matching entities with external databases}. Often, external databases include entities with varying levels of detail and granularity, making it rare to find an exact match.   Consequently, analysts usually settle for matches with similar or closely related, but not identical, entities to accomplish tasks at hand. {\bf Besides term fuzziness, a significant challenge in such an entity matching process lies in the ambiguity in defining what "match" means.} Understanding how entities are related and the nuances involved is especially important for high-stake tasks in responsible AI.
To illustrate this ambiguity, we start with an example problem based on our collaborators at ESG Flo:

\begin{figure}[ht]
  \centering
  \begin{subfigure}[b]{1\linewidth}
  \centering
    \includegraphics[width=0.8\linewidth]{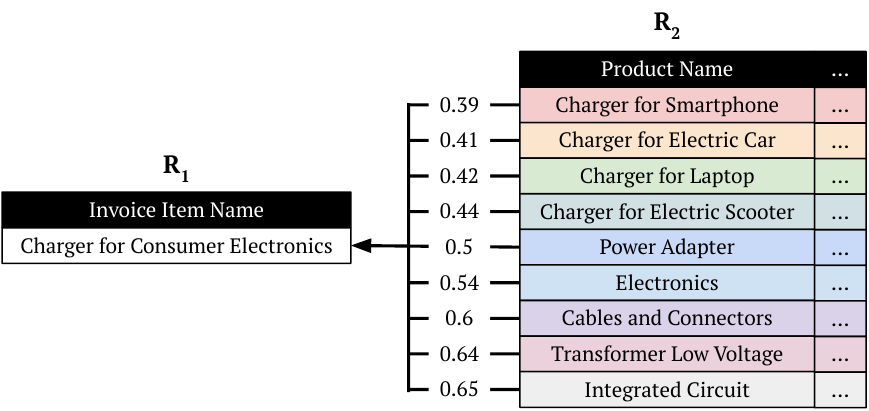}
    \caption{Using the {\it ada-002} Embedding and its Euclidean distance to rank the similarity of entities. However, many entities from table $R_2$ are related to those in $R_1$ in various ways that overwhelm analysts.}
    \label{fig:naive_em}
  \end{subfigure}
  
  \hfill 
  
  \begin{subfigure}[b]{1\linewidth}
    \centering
    \includegraphics[width=0.7\linewidth]{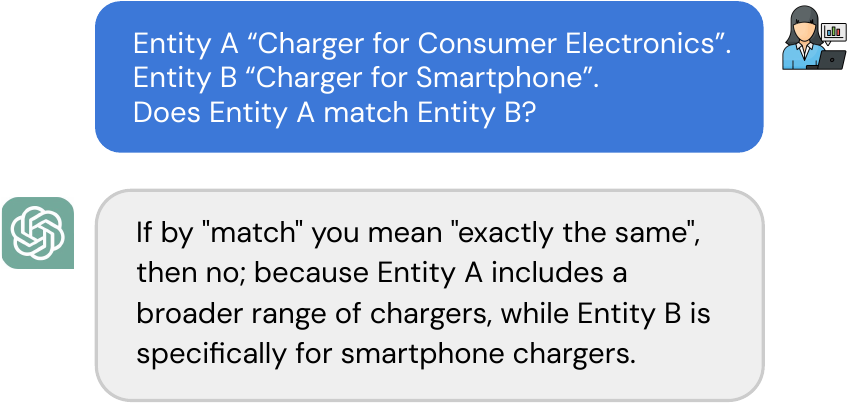}
    \caption{Using {\it GPT-4} for Entity Matching raises ambiguity in what "match" means. If "match" means "exactly the same", it is too strict to be practically useful as none of the entities are matched.}
    \label{fig:naive_em}
  \end{subfigure}

  \hfill 
  
  \begin{subfigure}[b]{1\linewidth}
    \includegraphics[width=0.9\linewidth]{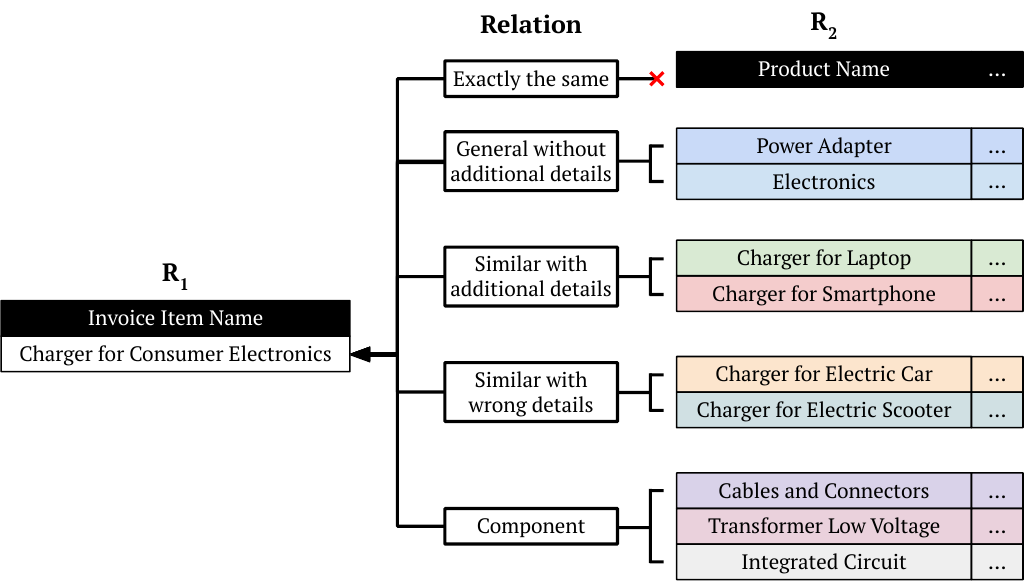}
    \caption{Discover the relations between source and target entities to disambiguate entity matching. Even if there are no entity with exactly the same, analysts settle for entities related in some predefined ways.}
    \label{fig:sub2}
  \end{subfigure}
  \caption{Entity Matching for ESG emission factor.}
  \label{fig:main}
\end{figure}

\begin{example}

ESG Flo is a startup that provides auditable data on environmental, social, and governance (ESG) factors. Their analysts work with a customer-provided table, $R_1$, consisting of item invoices. To assist customers in estimating carbon emissions for ESG reporting~\cite{epa_scopes_1_2_3}, the analysts aim to integrate $R_1$ with an external table, $R_2$, which provides ESG emissions factor estimates for various items. The challenge during data integration between $R_1$ and $R_2$ arises because the entities listed in $R_1$ and $R_2$ could be related but rarely identical. For example, consider the "Charger for Consumer Electronics" item in $R_1$ and the various issues that occur when naively using the current embedding and LLMs to match the entities in $R_2$:

\begin{itemize}[leftmargin=*,itemindent=0pt]
  
    \item {\bf Embedding:} When utilizing the ada-002 embedding to process all tuples in $R_1$ and $R_2$, and attempting to find semantically similar items for each tuple in $R_1$ by measuring Euclidean distance in the embedding space, analysts frequently encounter a broad array of items that seem related (as illustrated in \Cref{fig:naive_em}) but do not match exactly. They also find the Euclidean distance difficult to interpret.
    
    \item {\bf LLMs:} 
    When naively prompting LLMs, such as GPT-4, to determine whether two entities from $R_1$ and $R_2$ match, GPT-4 finds the term "match" ambiguous and has difficulty providing a confident answer. If "match" is interpreted as "exactly the same," this definition is too strict to be practically useful, as external databases rarely contain entities that are "exactly the same", and GPT-4 always answers "no".

\end{itemize}

Both Embedding and LLMs yield unsatisfactory results for the high-stakes task of ESG reporting. As a result, ESG Flo analysts previously relied solely on embedding to identify semantically similar entities, which then required considerable manual effort to understand their relationships and determine their relevance to the task. For the example input entity of "Charger for Consumer Electronics," they identified "Charger for Smartphone" as a more specific entity and "Power Adapter" as a more general one. After careful brainstorm, they prefer using "Power Adapter" for emission estimation because it is related but makes fewer assumptions, even though it is not an exact match and not the closest in the embedding space.
\end{example}

The issue of ambiguity is widespread in various entity matching tasks.
For instance, in the realm of ontologies such as OWL (Web Ontology Language)~\cite{mcguinness2004owl} and Schema.org~\cite{guha2016schema}, as well as in standardized vocabularies like the OHDSI Vocabularies~\cite{bathelt2021usage}, entities are often noisy and exhibit a lack of uniformity in terms of the levels of detail and granularity.
To effectively match entities, especially those used for high-stake tasks, it's necessary not only to identify semantically similar entities but also to understand their relationships to the input entities and the nuances in the difference, in order to identify the most appropriate one for the downstream task.

To responsibly identify the matched entities and clarify how they are matched, this paper proposes a novel approach by identifying a set of "relations" that analysts predefine as important for their task in LLMs. 
Our primary observation is that the entity matching process in practice is typically iterative, rather than a straightforward one-time process. Analysts often have a predefined list of relations relevant to their task in mind. They start by seeking entities that are "exactly the same", If such matches are not found, they may consider entities that are conceptually "similar but differ in details" for their estimation. Throughout the process, how these entities are "related“ is the crucial factor for decision-making:

\begin{example}
Continuing with the previous examples, while no exact match for the input entity "Charger for Consumer Electronics" is found, the analyst decides to explore alternative relations with target entities such as "general without additional entities", "similar with additional details", etc., as illustrated in \Cref{fig:main}. After identifying the relations, the analyst understands that "Power Adapter" represents a broad category without extra details, and "Charger for Smartphone" suggests a more specific category, implying the electronic item is a smartphone—an assumption the analyst prefers not to make. Meanwhile, entities like "Cables and Connectors" are components of the input entity but are considered too distant in relation. After careful consideration, the analyst chooses "Power Adapter" as the match for estimating carbon emissions for ESG reporting.
\end{example}

While the concept of relation has long been fundamental in understanding connections between entities and has been utilized in knowledge extraction~\cite{codd1970relational}, we are the first to apply it to facilitate disambiguation in entity matching and downstream tasks. These matched entities, the relations of how they are matched, and the thought process (potentially from LLMs) are provided in the subsequent human-in-the-loop (HIL) process to make the final decisions on how matched entities should be used for high-stakes tasks.

\section{Approach Overview}

In this section, we first formalize the relation-based entity matching. Then, we discuss the system design that incorporates relation-based entity matching for high-stakes tasks, such as ESG reporting.

\subsection{Problem Definition}

We follow the standard data model, where tables are denoted as $ R $, comprising a set of tuples $t$, and a list of attributes $ A $.
Given two tables $ R_1 $ and $ R_2 $, each entry is defined by attribute-value pairs. Traditional Entity Matching (EM) looks for a function $ f $:
$$ f: R_1 \times R_2 \rightarrow \{0, 1\}$$
This function $ f $ identifies if a pair $ (t_1, t_2) $ "matches", indicating they refer to exactly the same entity, by giving $ f(t_1, t_2) = 1 $; otherwise, it yields $ f(t_1, t_2) = 0 $.

However, in practice, finding "exactly the same matches" in external databases is rare, so users also seek entities that are related in different ways. With a domain of relation concepts $ REL $, {\bf relation-based EM} seeks a matching function $f$:
$$ f: REL \times R_1 \times R_2 \rightarrow \{0, 1\}$$
For every trio $ (rel, t_1, t_2) $ where $ rel \in REL $ denotes a type of relationship (such as "Is a", "Contains"), $ f(rel, t_1, t_2) = 1 $ if $ t_1 $ and $ t_2 $ are related by $ rel $, and $ f(rel, t_1, t_2) = 0 $ otherwise. The {\it relation-based EM} generalizes traditional EM,  as the "exactly the same" can be considered as just one of the many possible relations.

\stitle{Remark: } 
(1) Relations are not mutually exclusive. One entity can be related to another in multiple relations.
(2) Some relations, like "contains," are one-to-many. 
For \(\{t | (rel, t_1,t)=1 \land t \in R_2\}\), while each element is contained by \(t_1\), the set does not indicate a complete set of components. (3) When \(t_1\) is related to many entities for a given $rel$, how to choose the best match among them depends on the relation and may require a manual process. We consider it as a postprocessing step of relation-based EM.

\begin{figure}[htbp]
  \centering
  \includegraphics[width=0.8\linewidth]{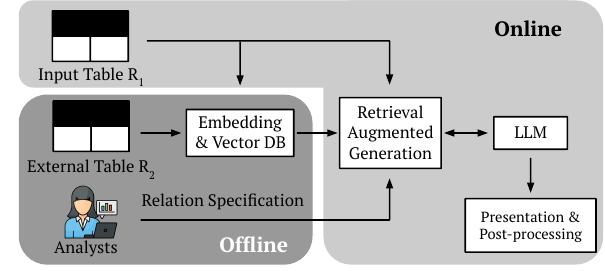} 
  \caption{System Design that performs relation-based entity matching for high-stake tasks like ESG reporting.}
  \label{fig:my_label}
\end{figure}

\subsection{System Design and Usage Walkthrough}

This section provides a walkthrough of the system design and its usage, which includes both offline and online phases. At a high level, we build embeddings offline for external tables to accelerate the online process. Online, we use LLMs to match entities based on a set of relations of interest. Finally, the matched entities, relations, and thoughts are presented for humans to verify for high-stakes tasks.

\subsubsection{Offline} Offline, analysts identify the external tables of entities that are of interest for the task, and preprocess the external table through embedding to accelerate online entity matching.

\stitle{Relation Specification.}
Analysts define a set of relations that are pertinent to the task during the offline brainstorming sessions. To identify these relations, analysts typically first perform entity matching manually.
Then, they analyze the patterns and common relations useful for the task.
The specificity of relation specifications is crucial, often enhanced by using examples in a few-shot learning context. This process is iterative, involving brainstorming and verification, and the relations may improve over time.

\begin{example}
For ESG Flo, after thorough discussions regarding the requirements, the analysts agreed upon the following relations:

\begin{itemize}[leftmargin=*,itemindent=0pt]
  
    \item \textbf{Exactly the same}: For the same entities, but with synonyms or abbreviations.
    E.g., "small automobile" is a synonym for "small car".
    
    \item \textbf{General without additional details}: For entities that are a general superclass of the input entity without additional details.
    E.g., "small vehicle" and "car" as general categories for "small car".
    
    \item \textbf{Similar with Additional Details}: For entities that include additional assumptions or features.
    E.g., "electric car" adds an assumption of electricity that "small car" does not imply.
    
    \item \textbf{Similar with wrong Details}: For similar entities but with details that contradict the input entity.
    E.g., "big car" contradicts the detail "small" in "small car" but they are both "car".
    
    \item \textbf{Component}: For parts or constituents of the given entity.
    E.g., "engine" as a component of a "small car".
    
\end{itemize}

\end{example}

\stitle{Embedding.} In the spirit of blocking~\cite{papadakis2020blocking}, embeddings are utilized to identify generally similar entities, but not related with respect to a specific relation as embedding is less interpretable. Given an external database for matching entities, we compute the embedding of the concatenation of attributes of its rows to expedite the discovery of relations during the online phase.
By default, the \texttt{ard-002} embedding model is employed, with the generated embeddings being stored in a \texttt{Faiss} index for efficient retrieval.

\subsubsection{Online Phase}
In the online phase, when provided with a user table, we perform entity matching, discover relations, and carry out post-processing according to the analysts' specific tasks.

\stitle{Retrieval Augmented Generation.}
We generate prompts that for each tuple $t \in R_1$ and each specified relation $rel$, identify the related tuples in $R_2$. Given $t$, we retrieve a set of $K$ entities from $R_2$ that are nearest $t$ in the embedding space. The prompt inquires, for each of the $K$ entities, whether it's related to $t$ by $rel$. We employ a standard chain-of-thought process to (1) enhance accuracy and assist in interpretation~\cite{wei2022chain} and (2) provide context for presentation during HIL. These prompts are then processed by a LLM (e.g., GPT-4). The following template is used:

\begin{Verbatim}[fontsize=\small, frame=single, framesep=2mm]
Task: Decide input & output entity relation.
Data: The input entity: {input_entity_row}
The output entities: {output_entity_rows}
Relation: {relation_description_with_example}
Steps:
1. Repeat input entity and relation.
2. Go through each output entity. 
Reason if it has the relation to input entity. 
Respond with JSON format:
{{
  "reasoning": "The input entity is ...",
  "matched entity indices": [0, ...],
  "explanation": "They are matched because..."
}}
\end{Verbatim}

The variable $K$ serves as a hyperparameter, balancing a trade-off: retrieving more entities improves recall but also increases computational demand and time. By default, we set $K = 10$. To improve recall, we continue to retrieve the next $K$ nearest entities in the embedding space if $\rho=30\%$ of the entities from the last batch are confirmed to have the specified relation.

\stitle{Presentation and Post Processing.}
Identifying the matched entities according to the specified relations is not the final step. For high-stakes tasks, user domain knowledge is needed to verify the matched entities and use them. After identifying the entities, we present (1) matched entities for different relations and (2) the explanations generated by LLMs to the analyst for post-processing. This step is iterative, beginning with matches that are "exactly the same". If a suitable match is not found, the process continues down a predefined list of relations for the next best estimation.

\begin{figure}
    \centering
    \includegraphics[width=0.9\linewidth]{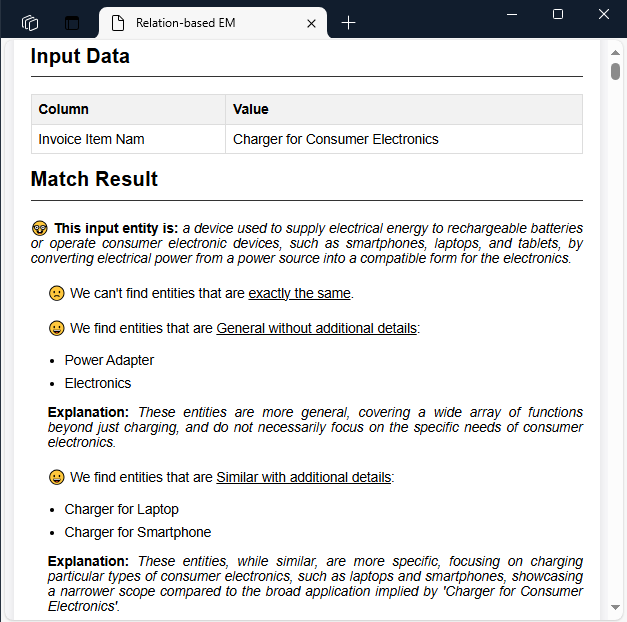}
    \vspace{-3mm}
    \caption{Generated report detailing the matched entities with respect to various relations, and their explanations, used by humans to perform downstream high-stakes tasks.}
    \vspace{-5mm}
    \label{fig:presentation}
\end{figure}

\begin{example}
For ESG Flo, the system generates a report, as illustrated in \Cref{fig:presentation}. Each entity is first checked for an "exactly the same" match. If found, this match is used.
If not, analysts try to use "General without additional details" as these are less assumptive. Among these, the most specific entity is selected.
Should entities with "Additional Details" be the next best match, these are then cross-referenced with customer data to ascertain the most suitable match.
In cases where "Similar with wrong Details" relations are found, efforts are made to identify the closest match.
Lastly, if the only match is "Component," the analysts aim to compile the most significant components and aggregate their data to estimate carbon emissions.
\end{example}

\section{User Study}

The system based on relation-based entity matching has been deployed at ESG Flo to facilitate the integration of external carbon emission databases for ESG reporting. We conducted a pilot study in which we interviewed an analyst using a think-aloud protocol to assess (1) the accuracy of the EM-matching report generated, and (2) the system's usefulness for the downstream ESG reporting task.

\stitle{Results.} The analyst finds the reports intuitive and useful for the reporting task: "The report that uses relations to match entities is very intuitive in its explanations. It aligns with how we thought about the entities when we manually generated ESG reporting. These reports have significantly reduced our manual efforts by saving us the effort of sifting through the candidates and allowing us to focus on quality verification to enhance the final reporting." However, the current systems have some insufficiencies due to their domain knowledge. "Since the reporting is for Category 1 [in the ESG framework], this entity is more general but is intended for capital goods, not purchased goods, which does not meet the requirement."
Addressing these requires a finer refinement of the relations and domain knowledge from models. This also underscores the importance of HIL design for these high-stakes designs.
Finally, the analyst points out a limitation of the current HIL design that may burden users: "The current number of input entity tables in the report is still a lot to review one by one. We find that input table tuples are similar, given that they're from the same companies. Most of these similar input tuples share the same relations with tuples in the external tables. It would be greatly beneficial if our manual verification for past tuples could be applied to the next similar tuples."
We plan to group the input tuples and discover relations for the batch to improve human effort in future work.

\section{Conclusion}
This work proposes a novel approach to entity matching that goes beyond traditional methods of identifying semantic similarities by emphasizing the importance of understanding and defining the relations between entities.  
By incorporating a HIL process and pre-defining a set of relevant relations, our approach enables analysts to more effectively navigate the spectrum of matched entities.

\begin{acks}
This work received funding from the Google PhD Fellowship. We thank the anonymous reviewers for their valuable feedback.
\end{acks}

\bibliographystyle{ACM-Reference-Format}
\bibliography{sample-base}

\appendix

\end{document}